\documentclass[aps,prl,twocolumn,superscriptaddress,groupedaddress,a4paper]{revtex4}
\usepackage{graphicx}
\usepackage{geometry}
\usepackage{color}

\begin{document}


\title{Comment on ``Non-Normalizable Densities in Strong Anomalous Diffusion:
Beyond the Central Limit Theorem''} 

\author{Igor Goychuk}
 
\affiliation{Institute for Physics and Astronomy, University of Potsdam, 
Karl-Liebknecht-Str. 24/25,
14476 Potsdam-Golm, Germany}

\begin{abstract}

A comment on the Letter by A. Rebenshtok, S. Denisov, P. H\"anggi, and E. Barkai,
Phys. Rev. Lett. \textbf{112}, 110601 (2014).

\end{abstract}

\maketitle

In a recent Letter \cite{Non}, an attempt is made to justify and
to provide a further support to earlier erroneous claims, see e. g. 
in \cite{Lutz}, that particle distributions or, equivalently, probability
densities can become non-normalizable in the case of anomalous L\'{e}vy walk
diffusion, and this anomaly can reflect a physical reality.
In this Comment, I show that  such claims cannot be justified and supported.

To begin with, any probability density function or PDF must be non-negative and normalized. 
Both a common
sense and the postulates of probability theory fix these important properties, once and forever,
see e.g. Ch. 2 in \cite{Papoulis}.  In Ref. \cite{Non}, Eq. (8) is derived 
from the correct result in Eq. (2) therein by taking  an 
asymptotic limit $t\to\infty$. The PDF in Eq. (2) is obviously
normalized for any particular model within the L\'{e}vy walk
model considered. The authors seem perfectly understand this,
see their Eq. (3).
However, already this fact
implies that Eq. (8) cannot be correct 
being applied to the \textit{whole} range of $x$ variations. 
Indeed, Eq. (8) describes a part 
of the distribution, for a sufficiently large $x$, however, not the 
\textit{entire} distribution. In this respect, Eq. (1) in 
\cite{Non} also describes only a tail of the time distribution $\psi(\tau)$. Clearly,
the distribution in Eq. (1) in \cite{Non} is non-normalizable, if by mistake to extend it 
to small $\tau\to 0$. However, to claim this for possible were absurd. An example
of PDF can be readily given. It is Pareto distribution,
$\psi(\tau)=\alpha/[\tau_0(1+\tau/\tau_0)^{1+\alpha}]$, where a constant $\tau_0$ separates 
small, $\tau\ll \tau_0$, and large, $\tau\gg \tau_0$, time intervals
between the scattering events. By the same token,
the claim that the distribution $P_A(x,t)$ in their Eq. (8) 
is not normalizable because 
$P_A(x,t)\sim |x|^{-1-\alpha}$ in the vicinity of $x\to 0$ is obviously wrong. 
The correct
distribution does not behave in this way for $|x|<v_0\tau_0$. 

After deriving incorrect result in Eq. (8), an attempt to further substantiate
it is made in \cite{Non} by using the concept of infinite covariant density (ICD).
Below I explain why this attempt fails. For this, let us clarify 
the origin of ICD in 
their Eq. (9). It is not a PDF,  
but a function concocted to be non-normalizable for  $\alpha>1$. 
Indeed, let us change random variable
from $x$ to $\bar v_{\alpha}=x/t^\alpha$, where $t$ is just a parameter, 
and not a random
variable. Then, the corresponding PDF is
\begin{eqnarray}
P_{v}^{(\alpha)}(\bar v_\alpha,t)=\int \delta(\bar v_\alpha-x/t^\alpha)P(x,t)dx\\ \nonumber
=t^{\alpha}
P(x=\bar v_\alpha t^\alpha,t)\;.
\end{eqnarray}
Most obviously, it is normalized for any  $t$, 
$\int P_{v}^{(\alpha)}(\bar v_{\alpha},t)d\bar v_{\alpha}=1$.
Hence, $P_\infty^{(\alpha)}(\bar v_\alpha):=\lim_{t\to\infty}P_v^{(\alpha)}(\bar v_\alpha,t)$
is also normalized. At the first look, our $P_\infty^{(\alpha)}(\bar v_\alpha)$ may seem
to be their ICD. However, this is not so. 
The function $I_{\rm cd}(\bar v)$ in Eq. (9) in \cite{Non} is not a
PDF of 
$\bar v \equiv \bar v_1$, which would be $\lim_{t\to\infty}P_v^{(1)}(\bar v_1,t)=
\lim_{t\to\infty} t P(x=\bar v t,t)=P_\infty^{(1)}(\bar v_1)$,  but a special construct. It can
be considered as the $t\to\infty$ limit of the function
$I_{\rm cd}(\bar v,t):=t^{\alpha-1}P_v^{(1)}(\bar v_1,t)$. Its normalization,
\begin{eqnarray}
\int I_{\rm cd}(\bar v,t)d \bar v= t^{\alpha-1},
\end{eqnarray}
indeed diverges for $\alpha>1$ in the limit $t\to\infty$. Of course, at any fixed and
finite observation time $t$, the (generalized) moment averages
\begin{eqnarray}\label{correct}
\langle |\bar v|^q\rangle&=&\frac{\int 
|\bar v|^q I_{\rm cd}(\bar v,t)d \bar v}{\int 
I_{\rm cd}(\bar v,t)d \bar v} \\
&=&\int |\bar v|^q  P_v^{(1)}(\bar v,t)d\bar v\nonumber\\
&=&t^{1-\alpha}\int |\bar v|^q I_{\rm cd}(\bar v,t)d \bar v\nonumber
\end{eqnarray}
can be found either with $I_{\rm cd}(\bar v,t)$, or with PDF 
$P_v^{(1)}(\bar v,t)$. However, it is incorrect to use
for this purpose $I_{\rm cd}(\bar v)$ obtained in the formal limit of infinite $t$,
as the authors do in their Eq. (13). It is generally, e.g. for $q<\alpha$, wrong.
The limit $t\to\infty$ must be taken in our Eq. (\ref{correct}).
Importantly, stochastic numerics can be done only at \textit{finite} 
$t$ and $\tau_0$. In this respect, which \textit{concrete} PDF
$\psi(\tau)$ was used in the numerics done in \cite{Non} is dim. 
Most obviously,
 any decent experiment, either real or numerical, done at \textit{finite} $t$ 
 will yield 
$I_{\rm cd}(\bar v,t)$, which is perfectly normalizable, and not $I_{\rm cd}(\bar v)$.
Hence ICD cannot correspond to a physical reality, contrary to the claims in \cite{Non}.

\end{document}